\documentclass[11pt,a4paper]{article}
\pdfoutput=1

\usepackage{jheppub}
\usepackage{latexsym}
\usepackage{multirow}
\usepackage{color}
\usepackage[usenames,dvipsnames,table]{xcolor}
\usepackage{graphicx}% Include figure files
\usepackage{epsfig}  % Include figure files
\usepackage{epsf}    % Include figure files
\usepackage{dcolumn}% Align table columns on decimal point
\usepackage{bm}% bold math
\usepackage{dcolumn}% Align table columns on decimal point
\usepackage{textcomp}% Align table columns on decimal point
\usepackage{float}
\usepackage{subfig}
\usepackage{hypcap}
\usepackage[]{hyperref}
\usepackage{makecell}
\usepackage{color}
\usepackage{pifont}
\usepackage{appendix}
  
\hypersetup{
  bookmarks=true,         % show bookmarks bar?
  unicode=false,          % non-Latin characters in Acrobat?s bookmarks
  pdftoolbar=true,        % show Acrobat?s toolbar?
 pdfmenubar=true,        % show Acrobat?s menu?
 pdffitwindow=true,     % window fit to page when opened
 pdfstartview={FitH},    % fits the width of the page to the window
 pdfsubject={Neutrino Oscillations Phenomenology},   % subject of the document
 pdfnewwindow=true,      % links in new window
 pdfcreator={RevTeX},
 colorlinks=true,       % false: boxed links; true: colored links
 linkcolor=red,          % color of internal links
 citecolor=blue,        % color of links to bibliography
 filecolor=black,      % color of file links
 urlcolor=blue,           % color of external links
  }

\makeatother
\title{Design, development and performance study of six--gap glass MRPC detectors}

\author[a,b]{Moon Moon Devi} 
\author[a]{, Naba K.~Mondal}
\author[a]{, B.~Satyanarayana}
\author[a]{, R.R.~Shinde} 

\affiliation[a]{Tata Institute of Fundamental Research, Mumbai 400005, India}
\affiliation[b]{Presently at: Weizmann Institute of Science, Rehovot 7610001, Israel }

\emailAdd{moon-moon.devi@weizmann.ac.il, devi.moonmoon@gmail.com}

\abstract
{
The Multigap Resistive Plate Chambers (MRPCs) are gas ionization detectors with multiple 
gas sub--gaps made of resistive electrodes. The high voltage (HV) is applied on 
the outer surfaces of outermost resistive plates only, while the interior plates are left 
electrically floating. The presence of multiple narrow sub--gaps with high electric field 
results in faster signals on the outer electrodes, thus improving the detector's time 
resolution. Due to their excellent performance and relatively low cost, the MRPC detector 
has found potential application in Time--of--Flight (TOF) systems. Here we present the 
design, fabrication, optimization of the operating parameters such as the HV, 
the gas mixture composition, and, performance of six--gap glass MRPC detectors of area 27cm $\times$ 
27 cm, which are developed in order to find application as trigger detectors, in TOF measurement etc. 
The design has been optimized with unique spacers and blockers to ensure a proper gas flow through 
the narrow sub--gaps, which are 250 $\mu$m wide. The gas mixture consisting of R134A, Isobutane and SF$_{6}$, and the 
fraction of each constituting gases has been optimized after studying the MRPC performance 
for a set of different concentrations. The counting efficiency of the MRPC is about 95\% at $17.9$ 
kV. At the same operating voltage, the time resolution, after correcting for the walk effect, is found 
to be about $219$ ps. 
}

\begin{document}
\maketitle
\flushbottom

%%%%%%%%%%%%%%%%%%%%%%%%%%%%%%%%%%%%%%%%%%%%%%%%%%%%%%%
\section{Introduction} 
\label{intro}

The Multigap Resistive Plate Chamber Detector was conceptualized and developed in 1996 \cite{CerronZeballos:1995iy, Williams:2001ms}.
These detectors are gas ionization detectors with multiple sub--gaps, consisting of resistive plates
(glass in our case) separated from one another using spacers of equal thickness. Even though there are many gaps, there is a
single set of anode and cathode readout electrodes, placed at the outer surfaces of the two outermost resistive plates.
 The interior plates are left electrically floating. The narrower sub--gaps enhance their time resolution capability. 
 
The results from groups involved in the study of various MRPC configurations indicate that a
time resolution of less than $100$ ps can easily be obtained with MRPC detectors. The MRPCs have been chosen as optimal elements for many
Time--Of--Flight (TOF) detector systems (including ALICE and STAR) due to their excellent time resolution
and higher efficiency for particle detection \cite{Fonte:1999cb}. The test prototype module with 4 gap MRPC detectors of area 3 m$^2$, developed for the very large area TOF array 
of the ALICE experiment, had a time resolution of about 70 ps \cite{Akindinov:2000rq, Williams:2001ms}. Later on, the MRPCs are upgraded 
with 5 sub--gaps each in two cells, and the time resolution was 50 ps or better \cite{Akindinov:2009zza, Alici:2013}. MRPC Modules of 6 gas--gaps of 0.220 mm and pad area of 20 cm$^2$, developed for 
the STAR experiment at RHIC, have a time resolution of the order of $60$ ps \cite{Bonner:2003, Wang:2005}. The MRPC detectors have also been tested in streamer and avalanche modes \cite{Narita:2009}.
The studies on the gas mixture also show that the baseline mixture of about 90\% of C$_2$F$_4$H$_2$, about 5\% of C$_4$H$_{10}$ and 
 the rest of SF$_6$ is suited for the detectors \cite{Akindinov:2004}.
 
 %%%%%%%%%%%%%%%%%%%%%%%%%%%%%%%%%%%%%%%%%%%%%%%%%%%%%%%%%%
\begin{figure}[t]
\centering
\includegraphics[trim=0cm 0.5cm 0cm 1cm, width=12cm]{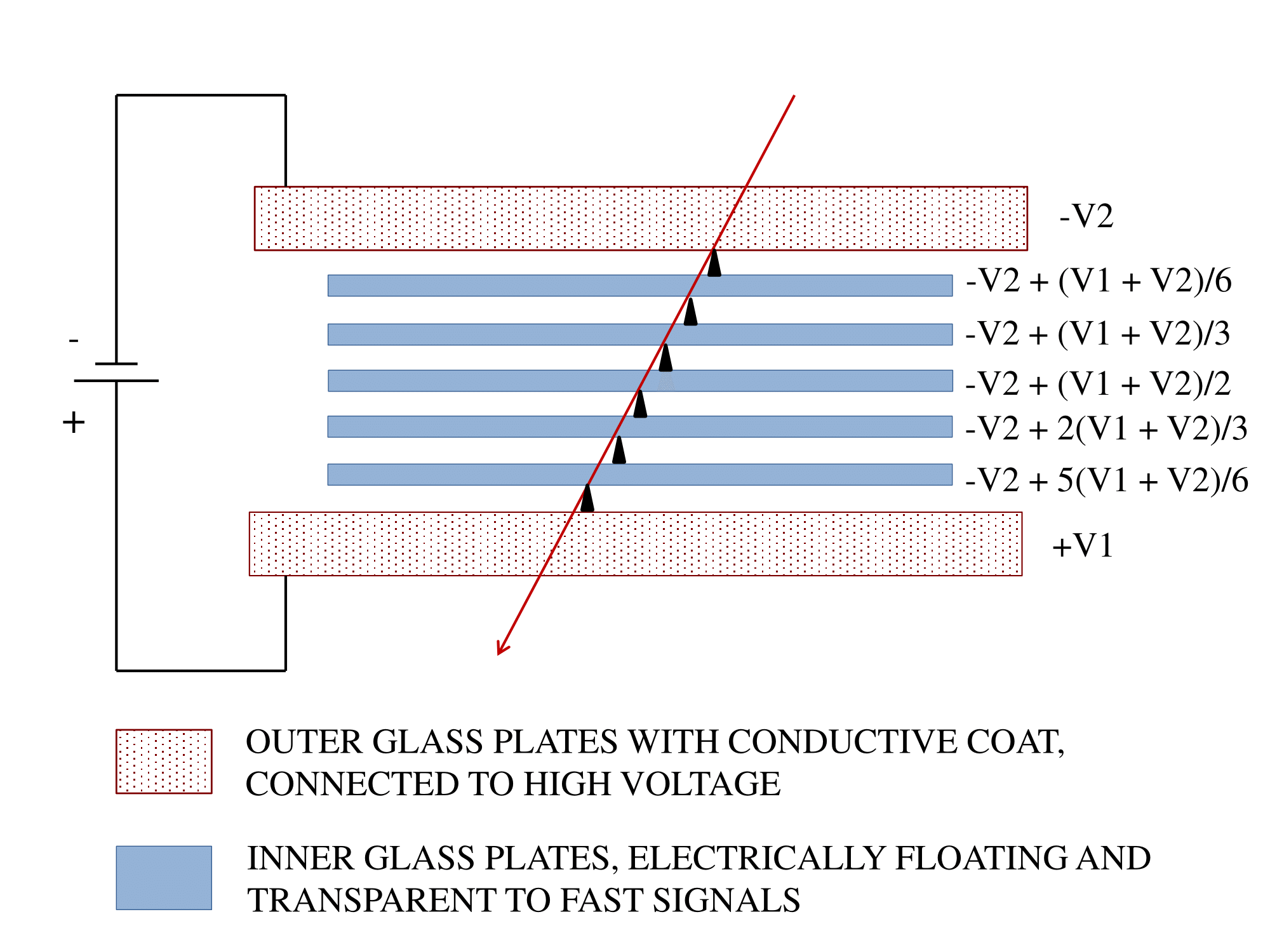}
\caption{The illustration of the potentials across the sub--gaps of an ideal MRPC detector.}
\label{mrpc_illus}
\end{figure}
%%%%%%%%%%%%%%%%%%%%%%%%%%%%%%%%%%%%%%%%%%%%%%%%%%%%%%%%%%

 The working principle, including the avalanche formation of an MRPC are similar to that of a single gap 
RPC, apart from the fact that, the additional sub--gaps improves the space--charge effect and limits the avalanche 
growth above a certain limit and reduces the time jitter of the signal.
An illustration of an MRPC in the ideal case is shown in Figure~\ref{mrpc_illus}, where all the gas gaps are
assumed to be of uniform width. The interior plates are electrically floating, and are maintained at equal voltages
due to the flow of positive ions and electrons between them. The voltage across each sub--gap is the same. Hence
on the average, each sub--gap will produce the same number of avalanches when a flux of charged particles
passes through it. This means the flow of electrons and ions into the plates bounding a gas gap will be identical
for all the gaps, and the net current to any of the internal plates would be zero. The surface resistivity of the 
conductive graphite coat is high enough so that the electrodes 
act as dielectrics, i.e., they are transparent to the fast signal 
generated by the avalanches inside each gas gap. The avalanche in any of the sub--gaps will 
induce a signal to the outermost electrodes, as 
the inner electrodes are transparent
to the fast signals. The fast signals in case of MRPC are produced by the flow of electrons towards the anode.
The resultant signal is the summation from all the gas gaps and it enhances the amplitude of the pulse. Copper
pickup strips placed outside the cathode and anode electrodes record the signal, with a reduced time jitter,
through induction. The intermediate plates act as the physical barriers to an excessive growth of the 
avalanche, and hence a
higher electric field can be applied to the detector operated in the avalanche mode, compared to that of a
single gap structure. This is advantageous in terms of the time resolution and rate capability of the device. The
strong uniform electric field stimulates the avalanche process immediately after the primary ionization is
created by a charged particle, leading to a good time resolution.

MRPCs may consist of a single stack with two external electrodes, or two
stacks packed together with three external electrodes, the anode being common
for both the stacks \cite{Williams:2001ms}. Single cell (stack) configuration has a pair of external electrodes.
As the number of floating electrodes increases, the operating voltage also increases. A double 
cell(stack) consists of two single cell MRPCs which are clamped together (usually the anodes). 
 The distribution of the floating electrodes in the
two cells reduces the operating voltage to be applied across each individual cell.

The India-based Neutrino Observatory (INO) \cite{bib5} is a proposed underground neutrino laboratory with a 
long--term goal of conducting decisive experiments in neutrino physics and will also house other experiments which
require an underground facility in future. Single gap RPC detectors have been chosen as the active 
detector elements for the magnetized Iron CALorimeter (ICAL) detector, due to
their high efficiency, position and timing characteristics besides their long-term suitability for large detector
coverage \cite{bib6}. However as a part of the extended R\&D at INO, MRPC detectors have been developed to find 
potential application in TOF detectors, medical imaging etc. 

Here we present the development and performance of single cell six--gap glass MRPC 
detectors with each sub--gap being about $250~\mu$m thick. In section~\ref{fabrication} we present the details 
of the fabrication procedure and the optimized design. The experimental setup, including the trigger and data acquisition
 system, are described in section~\ref{setup}. In section~\ref{results} we present the MRPC characteristics.
 
 \section{Fabrication}
\label{fabrication}

\begin{figure}[h]
\centering
\includegraphics[width=13cm]{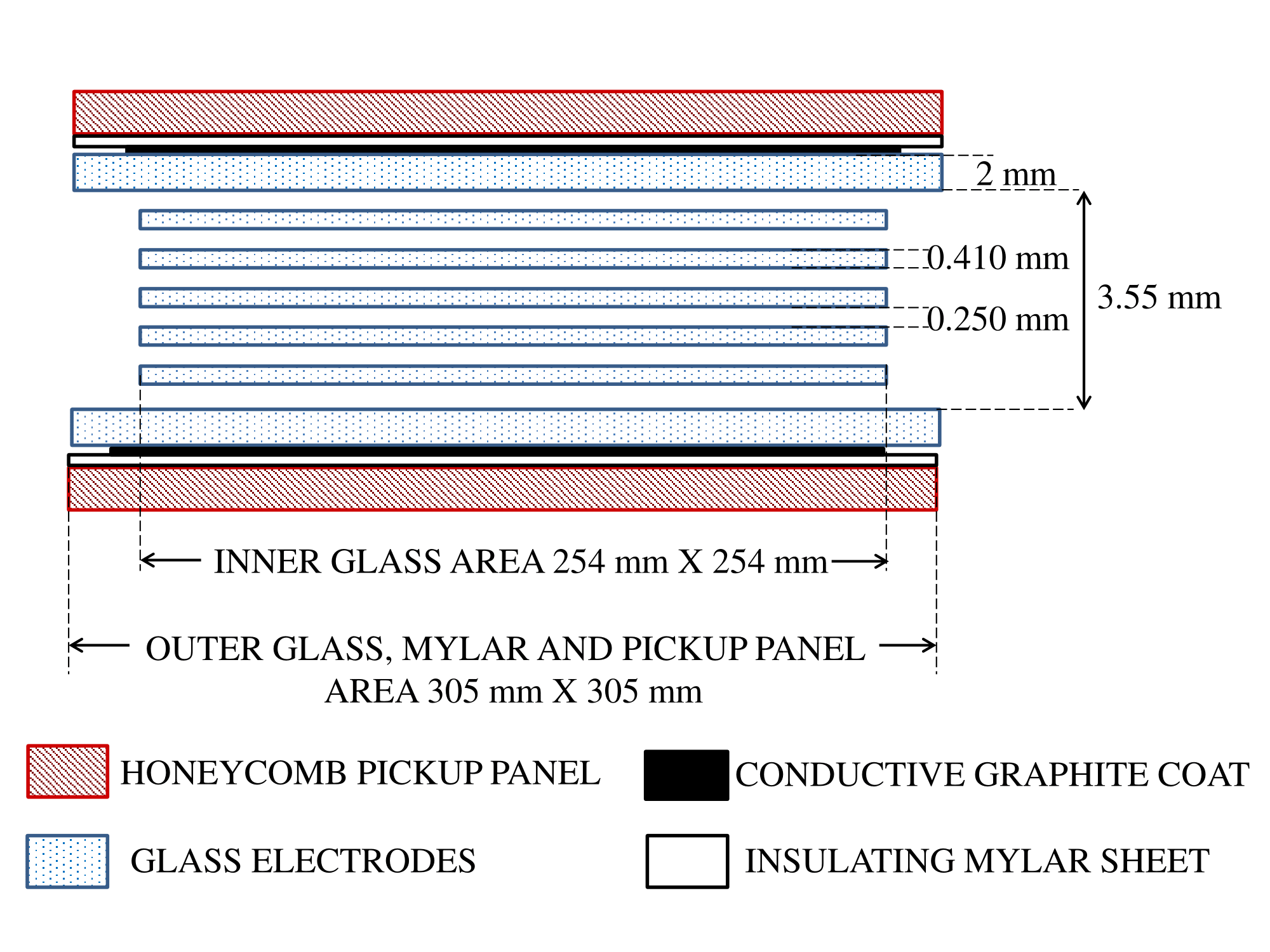}
\caption{The schematic (with dimensions) of the six--gap RPCs.}
\label{mrpc_schematics}
\end{figure}

In this section we shall discuss about the design and fabrication of the MRPC 
detectors. We have constructed six-gap 
glass MRPCs with single cell structure of dimensions $305$ mm $\times 305$ mm $\times 7.5$ mm. A schematic of the configuration 
with dimensions of various components is shown in Figure~\ref{mrpc_schematics}. Note that the area of the internal 
glass plates are of dimension $256$ mm $\times 256$ mm $\times 0.410$ mm. 

%%%%%%%%%%%%%%%%%%%%%%%%%%%%%%%%%%%%%%%%%%%%%%%%%%%%%%%%%%
%%%%%%%%%%%%%%%%%%%%%%%%%%%%%%%%%%%%%%%%%%%%%%%%%%%%%%%%%%%%%

%%%%%%%%%%%%%%%%%%%%%%%%%%%%%%%%%%%%%%%%%%%%%%%%%%%%%%%%%%
\begin{figure}[h]
\centering
\includegraphics[trim=0cm 0cm 0cm 0cm, width=10cm]{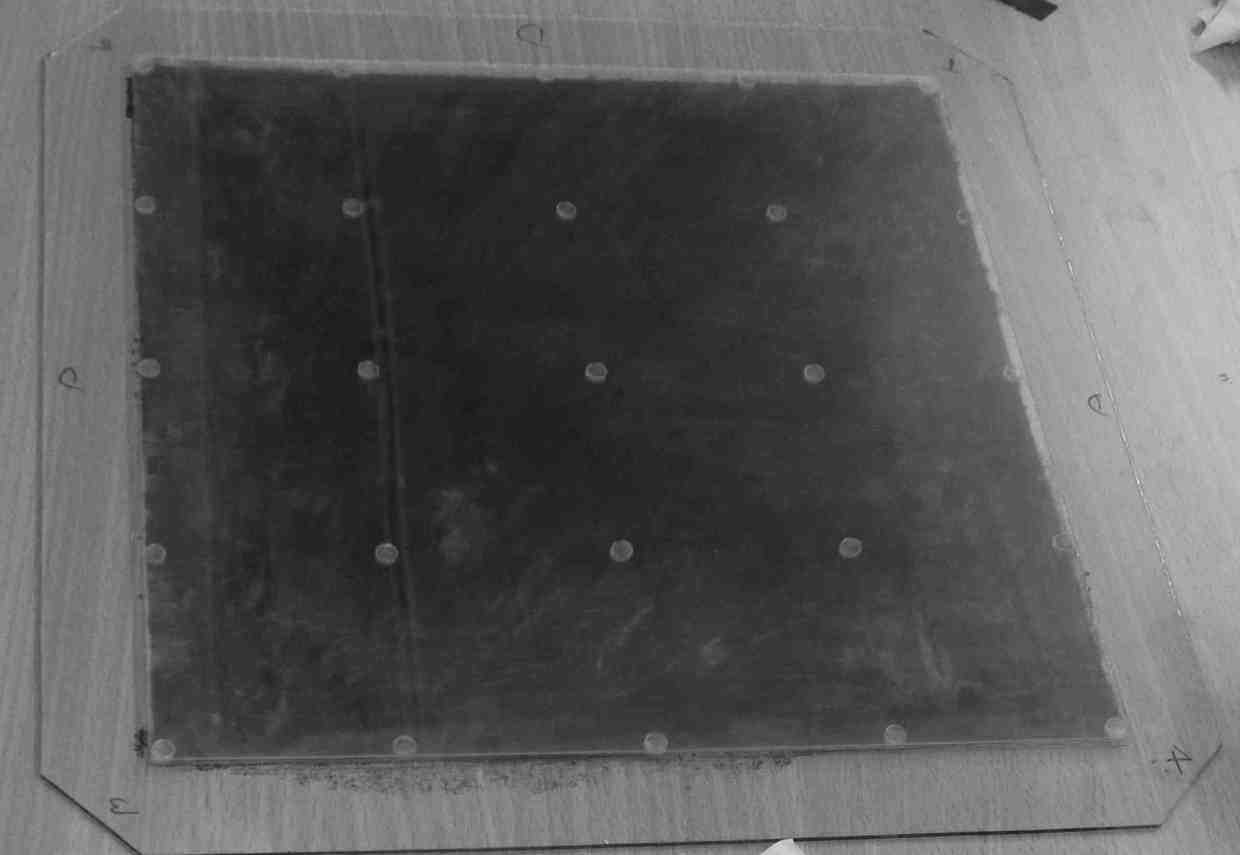}
\caption{The photograph of an MRPC showing the location of the spacers. There are $25$ 
spacers in each sub--gap in a $5~\times~5$ array, the
gap between any two consecutive spacers being $6.4$ cm.
}
\label{mrpc_spacer}
\end{figure}
%%%%%%%%%%%%%%%%%%%%%%%%%%%%%%%%%%%%%%%%%%%%%%%%%%%%%%%%%%

\begin{figure}[ht]
\centering
\includegraphics[trim=0cm 0cm 0cm 0cm,width=14cm]{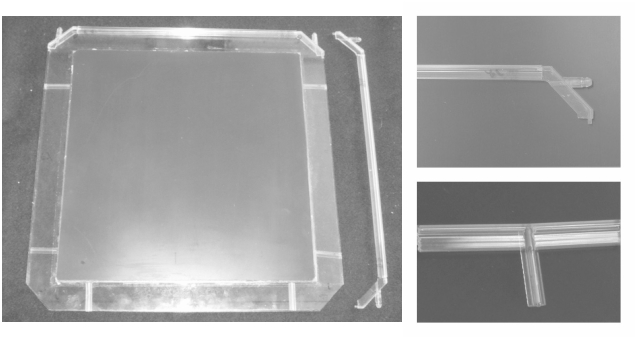}
\caption{\rm (Left) Photograph of an MRPC showing the placement of the blockers and side
spacer. Two blockers are placed near each gas inlet while 
one blocker each are placed near each gas outlet, 
to ensure a proper gas flow through the sub gaps. 
(Right--top) A side spacer fitted with a gas nozzle. 
(Right--bottom) A segment of a side spacer with the blocker attached.
}
\label{mrpc_blocker}
\end{figure}

Glass sheets of $2$ mm thickness, coated with a conductive layer using graphite and paint of 
the NEROLAC brand, were used for the outer electrodes.
 The surface resistance of the conductive 
coat was in the range ($0.5$ -- $1$) M$\Omega$/$\square$. 
 Two sided non 
conducting adhesive tapes were used on both sides of a mylar sheet to make small circular 
spacers of diameter $4$ mm and thickness $250~\mu$m. Twenty five spacers were used to maintain each gas gap. 
The placement of the 
spacers were shown in Figure~\ref{mrpc_spacer}. First
a few trials were made by placing this configuration in an enclosed box filled with the gas mixture. Such an
 enclosed structure had some drawbacks such as difficulty in alignment and the problem 
of ensuring sufficient and uniform gas flow through the sub--gaps. 
The configuration was optimized by sealing the gas gaps
using side spacers glued between the outermost electrodes.  As can be seen in 
figures \ref{mrpc_schematics} and \ref{mrpc_spacer}, there is a gap of
around $2.7$ cm from the edges of the external electrodes to the edges of the
internal electrodes. There is a possibility of gas following that path of thickness 3.55 mm, instead of
flowing through the 0.250 mm thick sub--gaps which would offer much higher resistance to the gas flow. In order to
ensure a proper flow through the sub--gaps, we introduced some blockers
at appropriate places (one each near the gas inlets and two each near the gas
outlets). This is illustrated in Figure~\ref{mrpc_blocker}. The pickup panel consists 
of honeycomb panels laminated with copper strips of width $2.8$ cm. The pickup strips on
both the sides of an MRPC were placed parallel to each other.

%%%%%%%%%%%%%%%%%%%%%%%%%%%%%%%%%%%%%%%%%%%%%%%%%%%%%%%%%%

\section{The Experimental Set-up}
\label{setup}
%\newpage
%%%%%%%%%%%%%%%%%%%%%%%%%%%%%%%%%%%%%%%%%%%%%%%%%%%%%%%%%%
\begin{figure}[t]
\centering
\includegraphics[trim=0cm 0cm 0cm 0cm, width=12cm]{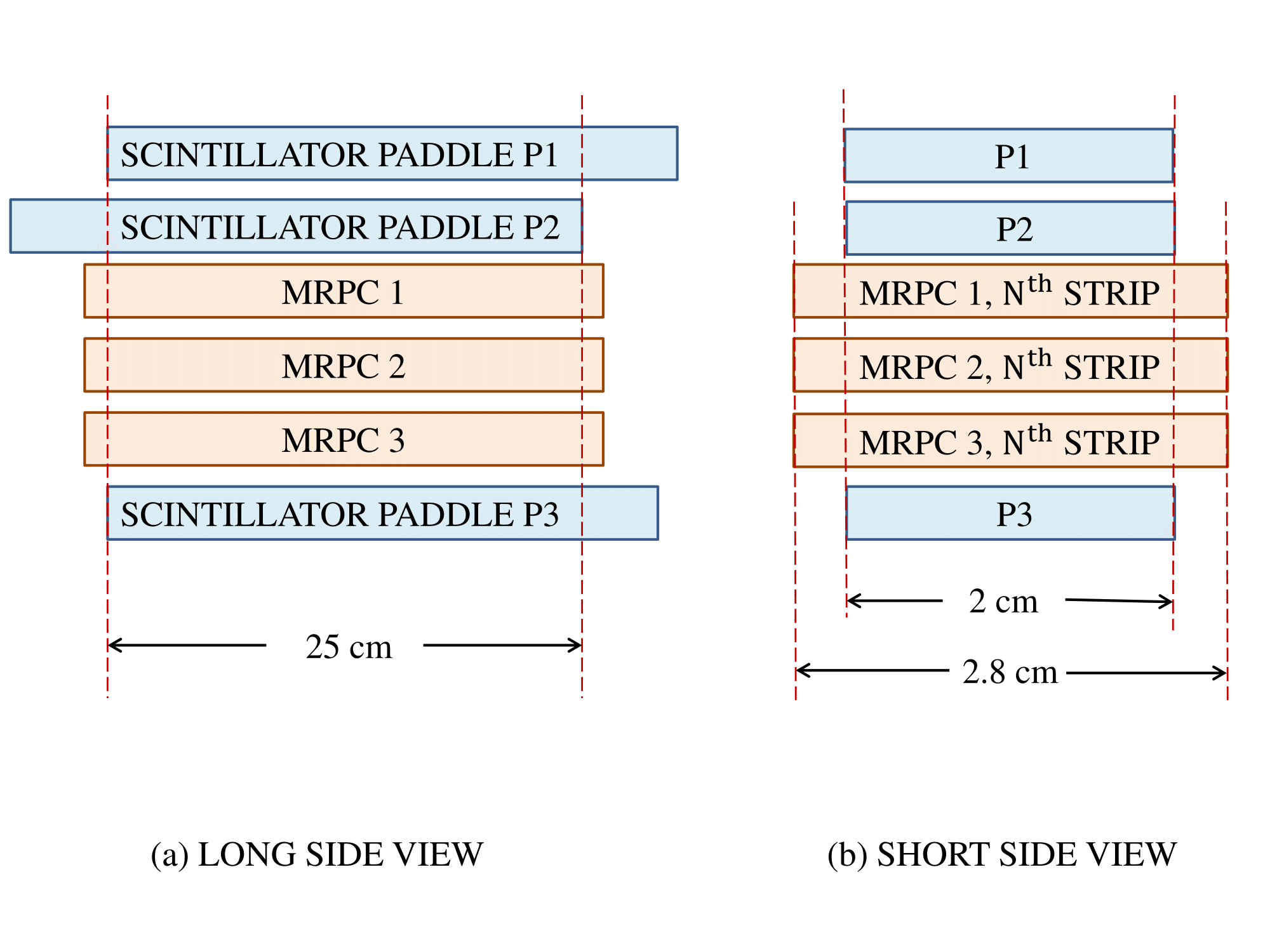}
\caption{Schematic for the cosmic ray muon telescope. P1, P2 and P3 
are scintillator 
paddles of width $2$ cm each, and they are aligned on a pick-up 
strip of width $2.8$ cm. The effective area of this telescope is $25$ 
cm $\times 2$ cm.}
\label{mrpc_trigger}
\end{figure}
%%%%%%%%%%%%%%%%%%%%%%%%%%%%%%%%%%%%%%%%%%%%%%%%%%%%%%%%%% 

The experimental set--up to test three MRPCs with the optimized design is described here. 
A cosmic muon telescope consisting of three scintillator paddles has been set--up. The details of the 
telescope, the preamplifier and the data acquisition system are described in the following subsections.
\subsection{The cosmic muon telescope}
\label{telescope}
\begin{figure}[h]
\centering
\includegraphics[width=12cm]{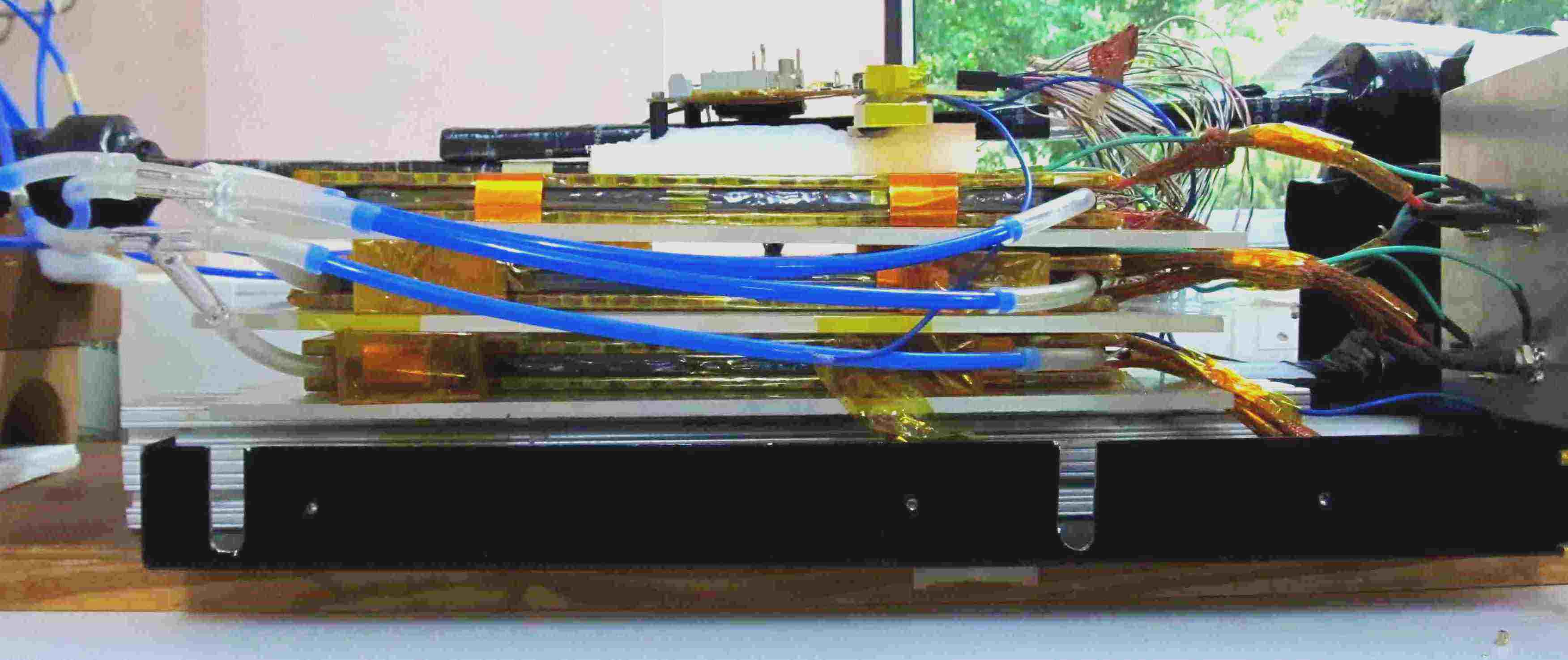}
\caption{The experimental set--up.}
\label{mrpc_setup}
\end{figure}

The MRPCs were operated in the avalanche mode and characterized using cosmic muons. Three
scintillator paddles of width $2$ cm each (two on the top of the MRPCs under test and one at the bottom) were set up
to construct a cosmic ray muon telescope as illustrated in Figure~\ref{mrpc_trigger}. A fast time coincidence of signals from these paddles indicates
the passage of a cosmic ray muon through the detector set--up. This coincidence signal has been used
 to trigger the data acquisition system. The set--up including three MRPCs and three scintillator paddles is shown in Figure~\ref{mrpc_setup}.

%%%%%%%%%%%%%%%%%%%%%%%%%%%%%%%%%%%%%%%%%%%%%%%%%%%%%%%%%%
%%%%%%%%%%%%%%%%%%%%%%%%%%%%%%%%%%%%%%%%%%%%%%%%%%%%%%%%%%

\begin{table}
\centering
\caption{Counting rate of an MRPC strip at different NINO thresholds.}
\label{table_nino}
\begin{tabular*}{\columnwidth}{@{\extracolsep{\fill}}cc|cc@{}}
\hline
%\multicolumn{2}{@{}l}{parameter} & Set 1 & Set 2\\
%Equil. & \multicolumn{1}{c}{$x$} & \multicolumn{1}{c}{$y$} & \multicolumn{1}{c}{$z$} & \multicolumn{1}{c}{$C$} & S 
Threshold~voltage & Count~rate& Threshold~voltage &\rm Count~rate\\
(mV) & (Hz)& (mV)&\rm (Hz)\\
\hline
$310.4$&$27$&$157$&$134$\\
\hline
$220.1$&$32$&$151$&$316$\\
\hline
$190.2$&$41$&$149$&$621$\\
\hline
$181.8$&$46$&$115.3$&$954$\\
\hline
\end{tabular*}
\end{table}

\subsection{NINO ASIC}
\label{nino}
For amplification and digitization, NINO ASIC, an ultra fast front end 
preamplifier-discriminator chip which was developed for the 
ALICE TOF experiment, was used \cite{bib7}. Each chip has $8$ amplifier and discriminator channels. Each channel is
designed with an amplifier with \textless~$1$ ns peaking time, a discriminator with a
minimum detection threshold of $10$~fC, and an output stage.  Each channel in the 
NINO ASIC chip takes
differential signals from the pickup strips as input, and
amplifies them in a four stage cascade amplifier.
The threshold to the discriminator stage of the chip was
set at $157$ mV after studying count rates of the
detector at various values as summarized in Table~\ref{table_nino}.

\subsection{DAQ}
\label{daq}
%%%%%%%%%%%%%%%%%%%%%%%%%%%%%%%%%%%%%%%%%%%%%%%%%%%%%%%%%% 

\begin{figure}[ht]
\centering
\includegraphics[trim=0cm 0cm 0cm 0cm, width=13cm]{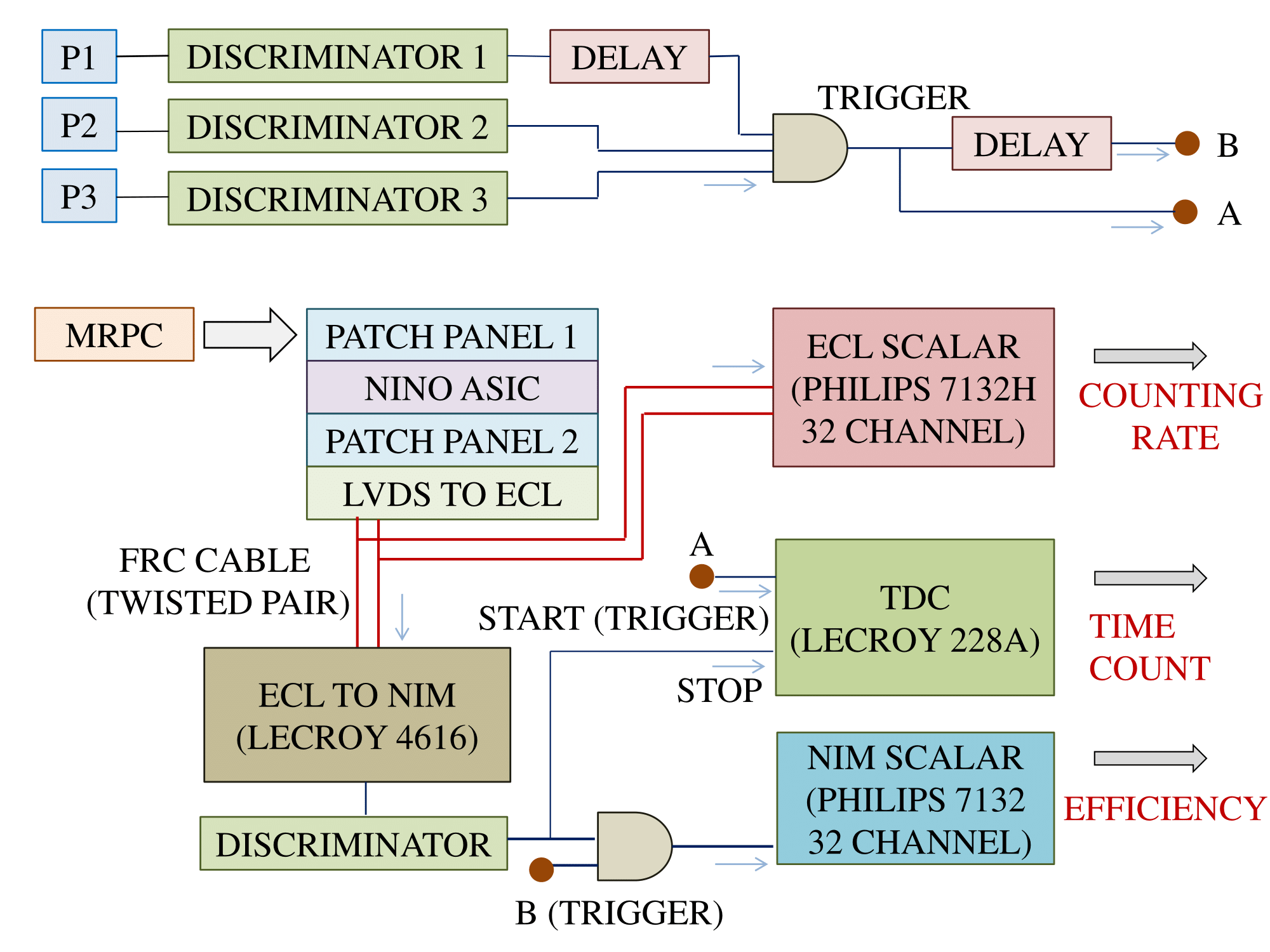}
\caption{The DAQ scheme to obtain the strip counting rate, efficiency and 
time count.}
\label{daq1}
\end{figure}
%%%%%%%%%%%%%%%%%%%%%%%%%%%%%%%%%%%%%%%%%%%%%%%%%%%%%%%%%% 

We have assembled a CAMAC based data acquisition (DAQ) system 
for the MRPC detector test set--up. MRPC pickup strips are amplified 
and digitized by NINO ASICs and then taken to the time coincidence 
unit for generating the trigger. The differential (LVDS) signals obtained from NINO outout 
are converted to ECL and then according to the requirement
of the scalar and TDC used directly or further converted to NIM signals. The 
counting rate of the individual strips are recorded with a ECL scalar.  
The trigger (T) is formed by generating a coincidence between the three scintillator paddles 
P$_1$, P$_2$ and P$_3$. The efficiency of the MRPC strips 
is then obtained from the coincidence of the trigger and the strip count. 
Note that, for parts of the time resolution study the NINO ASIC has been replaced 
by ANUSPARSH \cite{anusparsh}, a front--end ASIC designed for the INO-ICAL experiment, which also provides 
the analog output signal.
\begin{equation}
\rm Strip~efficiency = \frac{\rm MRPC~strip~count}{\rm T}
\end{equation}

The trigger is also given as the start to the TDC module to get the time count, 
where the Stop signal comes from the MRPC strips.
The DAQ scheme to obtain the counting rate, efficiency and timing is shown in 
Figure~\ref{daq1}.

\section{The MRPC performance}
\label{results}
Here we discuss the characteristics of MRPC obtained by adjusting 
various parameters such as gas mixture composition, HV etc. The 
I-V characteristics, strip count rate, efficiency and the time resolution 
has been studied.

\subsection{MRPC Characteristics as a function of Gas mixture and HV} 
\label{gas-hv}

%%%%%%%%%%%%%%%%%%%%%%%%%%%%%%%%%%%%%%%%%%%%%%%%%%%%%%%%%
%%%%%%%%%%%%%%%%%%%%%%%%%%%%%%%%%%%%%%%%%%%%%%%%%%%%%%%%%%
\begin{figure}[ht]
\centering
\includegraphics[trim=0cm 0cm 0cm 0cm, width=13cm]{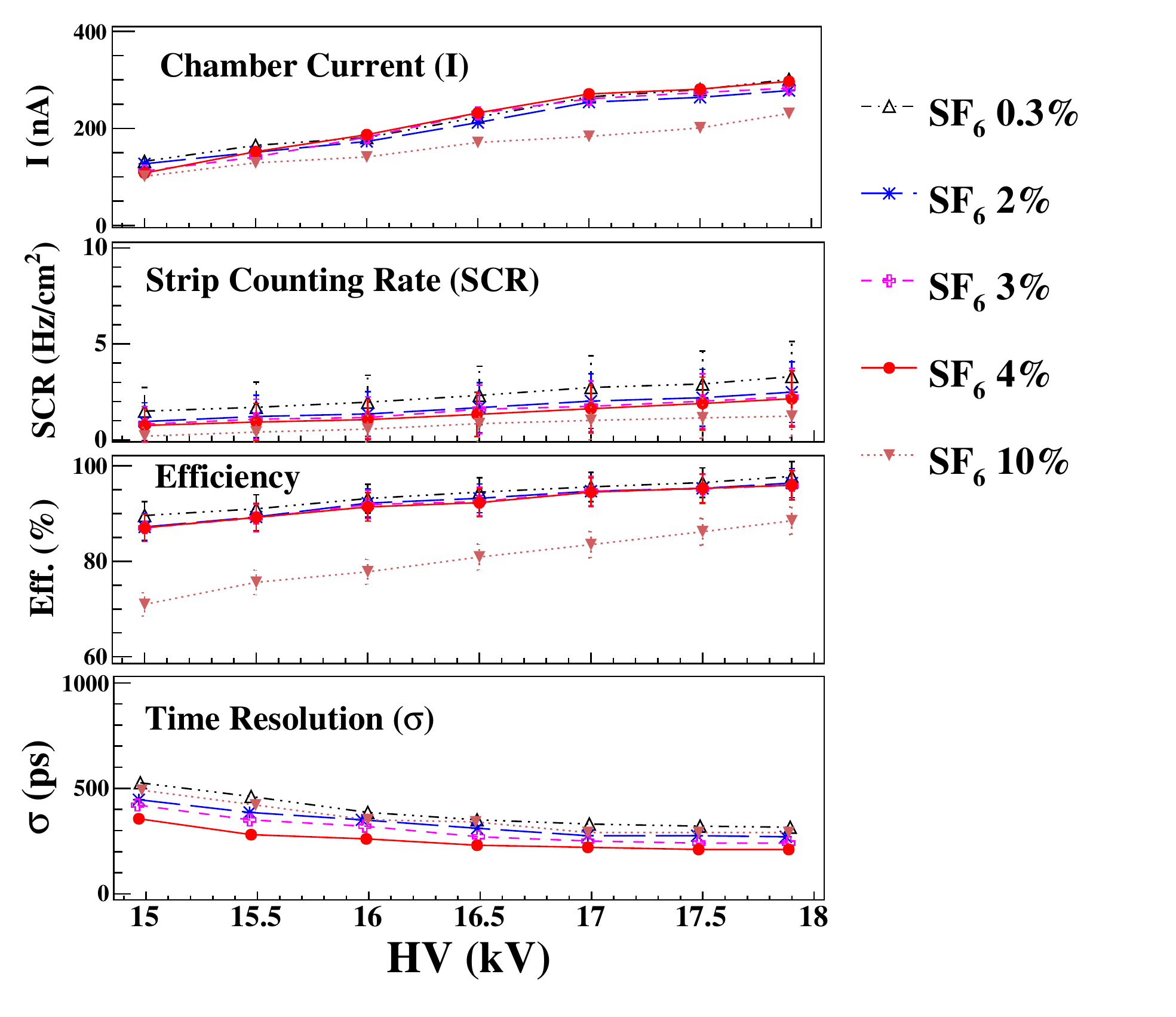}
\caption{The various characteristics of an MRPC strip as a function 
of the high voltage applied across the gas gap at different concentrations 
of the gas mixture of R134A, C$_4$H$_{10}$ and SF$_6$.}
\label{hv-test}
\end{figure}
%%%%%%%%%%%%%%%%%%%%%%%%%%%%%%%%%%%%%%%%%%%%%%%%%%%%%%%%%

The gas mixture is composed of R134A, C$_4$H$_{10}$  and SF$_6$ gases. Various studies show that 
the MRPCs are best operated with $\sim$ $5\%$ of SF$_6$ unlike the standard composition used to 
operate single gap RPC. With increasing SF$_6$ fraction, two competing processes affect the MRPC 
characteristics\cite{Akindinov:gasmixture}. Higher electric fields are required with increasing fractions of SF$_6$, which 
also increases the drift velocity and results in an improved time resolution. On the contrary, 
since SF$_6$ has large capture cross sections for low energy electrons, increasing the SF$_6$ concentration 
reduces the growth in the avalanche significantly. This leads to a reduction in 
the MRPC counting rate and efficiency, and a worsening in the time resolution.
So, an optimization of the gas mixture composition is required for balancing these two opposing effects.

%\newpage
  We have performed a study with various concentrations of SF$_6$ at different applied high 
voltages to 
obtain an optimized set. For this, the proportion of C$_4$H$_{10}$  was kept fixed at $5\%$, and the 
other two were varied. Figure~\ref{hv-test} shows the efficiency, counting rate per area of the pick-up strip, the 
chamber current and the time resolution of an MRPC strip as a function of the applied HV which was varied 
between $15$ kV and $17.9$ kV. We see that at $\sim~4\%$ of 
SF$_6$, the time resolution is the best and the noise rate and chamber current are reasonable 
without deteriorating the efficiency. So for further 
study, we have used the gas mixture of R134A ($91\%$), C$_4$H$_{10} (5\%)$
 and SF$_6 (4\%)$. We see that even at $17.9$ kV 
the chamber current and counting rate are not too high. 
In Figure~\ref{hv-test}, the time resolution of the MRPC, after correcting for the time 
walk has been shown. This correction is done via a calibration of the prompt time peak with 
the total charge deposited in an event. This is described in the following section.

\subsection{Time resolution} 
\label{time-reso}

%%%%%%%%%%%%%%%%%%%%%%%%%%%%%%%%%%%%%%%%%%%%%%%%%%%%%%%%%%
\begin{figure}[ht]
\centering
\includegraphics[trim=0cm 0cm 0cm -0.6cm, width=12cm]{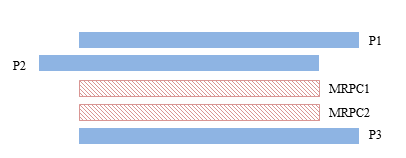}
\caption{\rm The set-up of two MRPCs and a trigger from the coincidence of three 
scintillator paddles. The scintillators are 2 cm wide and aligned on two chosen pick-up strips, one each 
from MRPC1 and MRPC2.}
\label{tmng}
\end{figure}
%%%%%%%%%%%%%%%%%%%%%%%%%%%%%%%%%%%%%%%%%%%%%%%%%%%%%%%%%%

\begin{figure}[ht]
\centering
\includegraphics[trim=0cm 0cm 0cm -0.6cm, width=12cm]{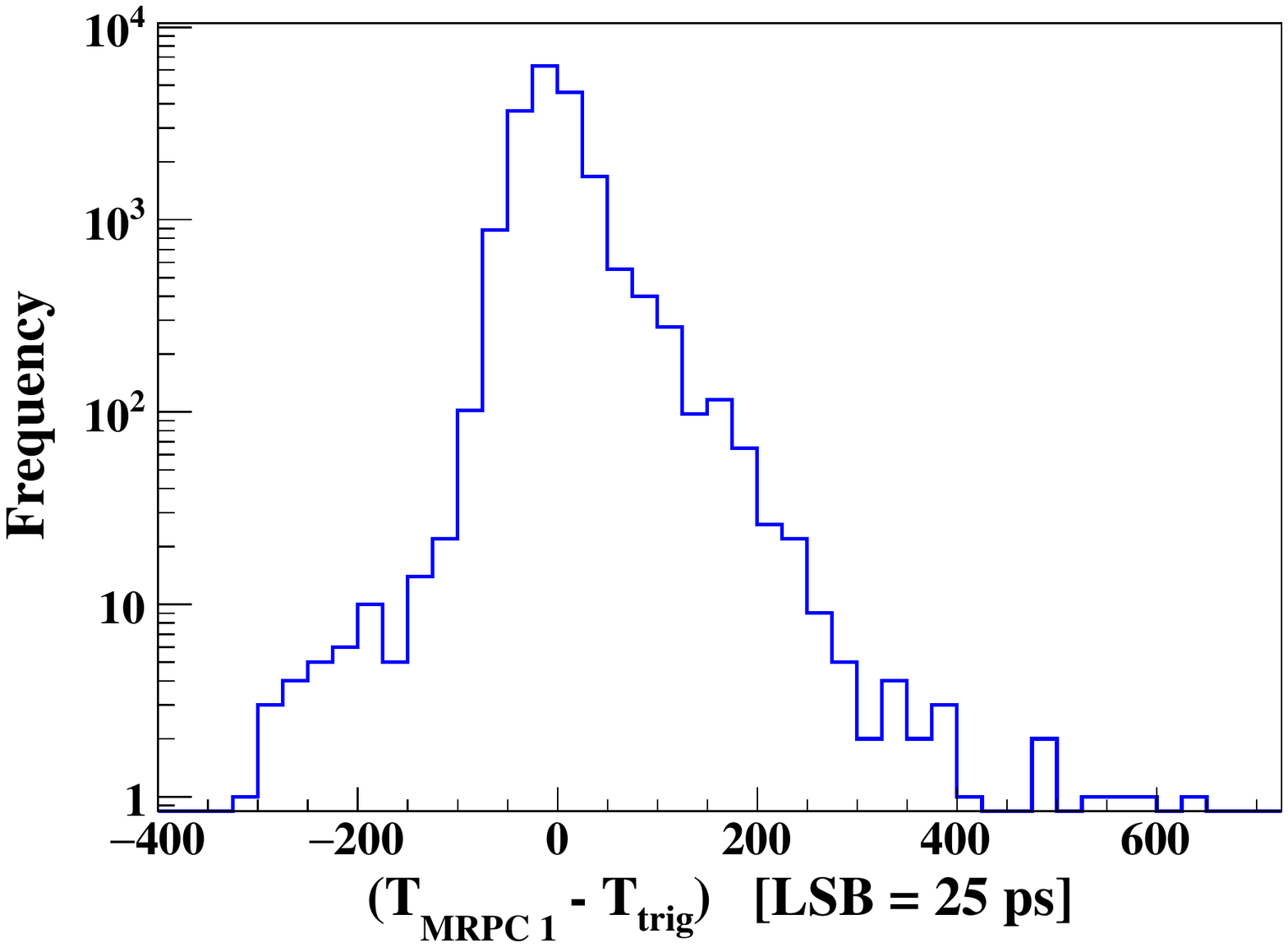}
\caption{\rm The raw MRPC time distribution with respect to the trigger at $17.9$ kV and 
with the gas mixture R134A ($91\%$), C$_4$H$_{10}$ $5\%$ and SF$_6$ ($4\%$). Note that the distribution has been 
shifted so that its peak is at zero.}
\label{tdc-raw}
\end{figure}

The introduction of smaller gas sub--gaps results in an 
improved time resolution in MRPCs. 
For an accurate measurement of the timing, it is important to 
reduce any fluctuations which may occur during the generation of the timing logic signal.  A major 
source of finite time resolution is the {\it walk} effect. This effect is caused by the variation in the 
signal amplitudes and/or rise time. The signals with different amplitudes 
cross the discriminator threshold at different times, resulting in a time shift (walk) in the logic signal. 
An additional walk effect arises due to the finite amount of charge that is required to 
be integrated on a capacitor to 
trigger the discriminator. We reduce the time walk by calibrating the time counts with the charge deposited. 
A setup of two MRPCs and three scintillator paddles, as shown in Figure~\ref{tmng}, has been used for this study. 
The trigger has been provided by the coincidence of the three scintillator paddles 
and this has been used as the `Start' to the TDC. A delay is added 
to the scintillator P1, so that the timing of the trigger is governed by it. 
In Figure~\ref{tdc-raw}, a distribution 
of the MRPC timing, with respect to the scintillator trigger has been shown.
A correction for time walk has been made 
according to the charge of the signal, which is described in the following. Note that, the charge is obtained from 
the analog output signal from the MRPC strips. Since the NINO ASIC provides 
LVDS output only, for this study it has been replaced with ANUSPARSH, a front-end 
ASIC designed for the INO-ICAL experiment.

%%%%%%%%%%%%%%%%%%%%%%%%%%%%%%%%%%%%%%%%%%%%%%%%%%%%%%%%%%
%%%%%%%%%%%%%%%%%%%%%%%%%%%%%%%%%%%%%%%%%%%%%%%%%%%%%%%%%%

\begin{figure}[t]
\centering
\includegraphics[trim=0cm 0cm 0cm 0cm, width=11cm]{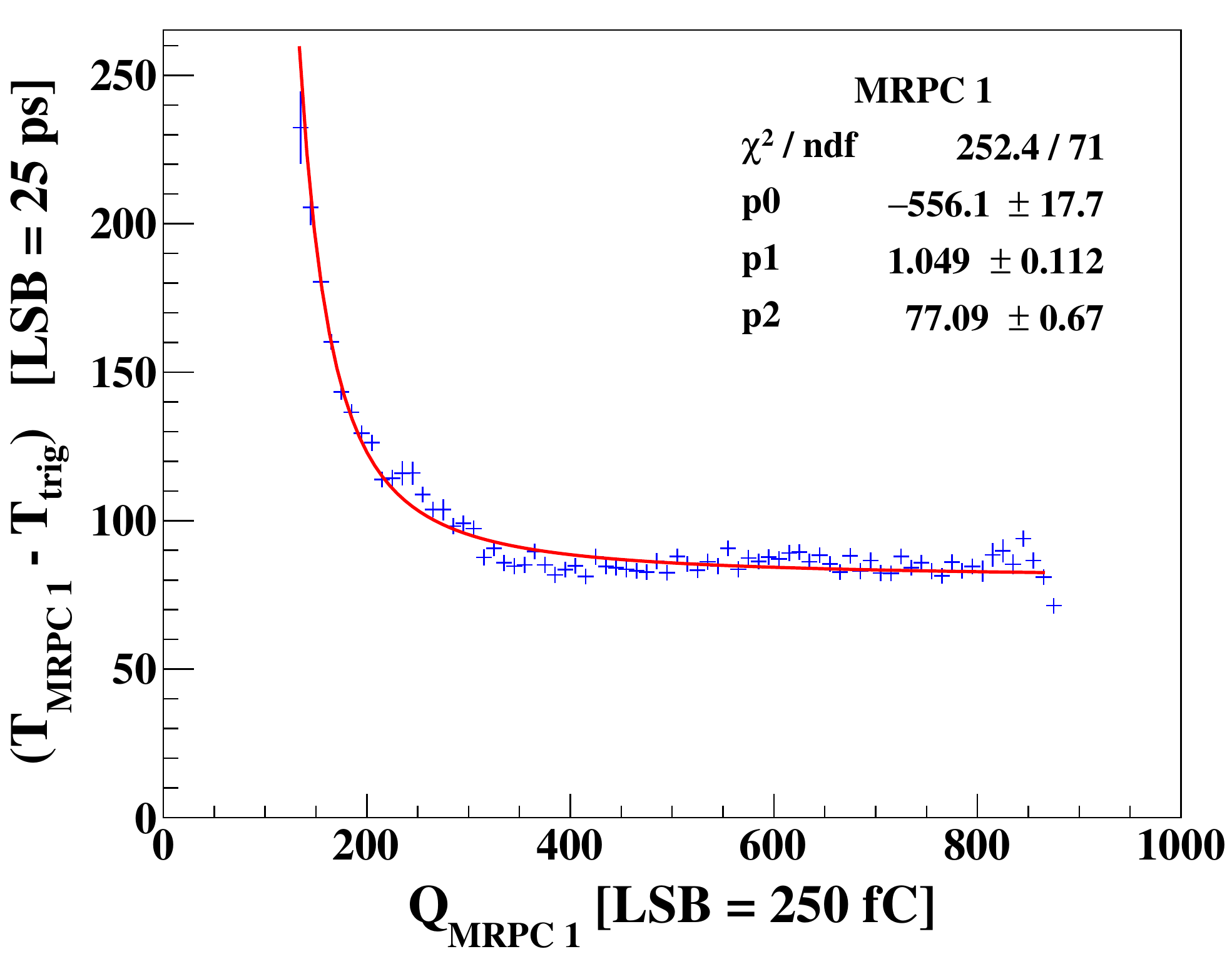}
\caption{ The calibration graph for correcting of the 
MRPC time distribution for time-walk, fitted to $\exp[- p_0/x + p_1 ] + p_2$. It is a profile 
histogram that shows the mean and rms of the scatter in each time count vs charge count bin.}
\label{calibration}
\end{figure}

\begin{figure}[h]
\centering
\includegraphics[trim=0cm 0cm 0cm 0cm, width=11cm]{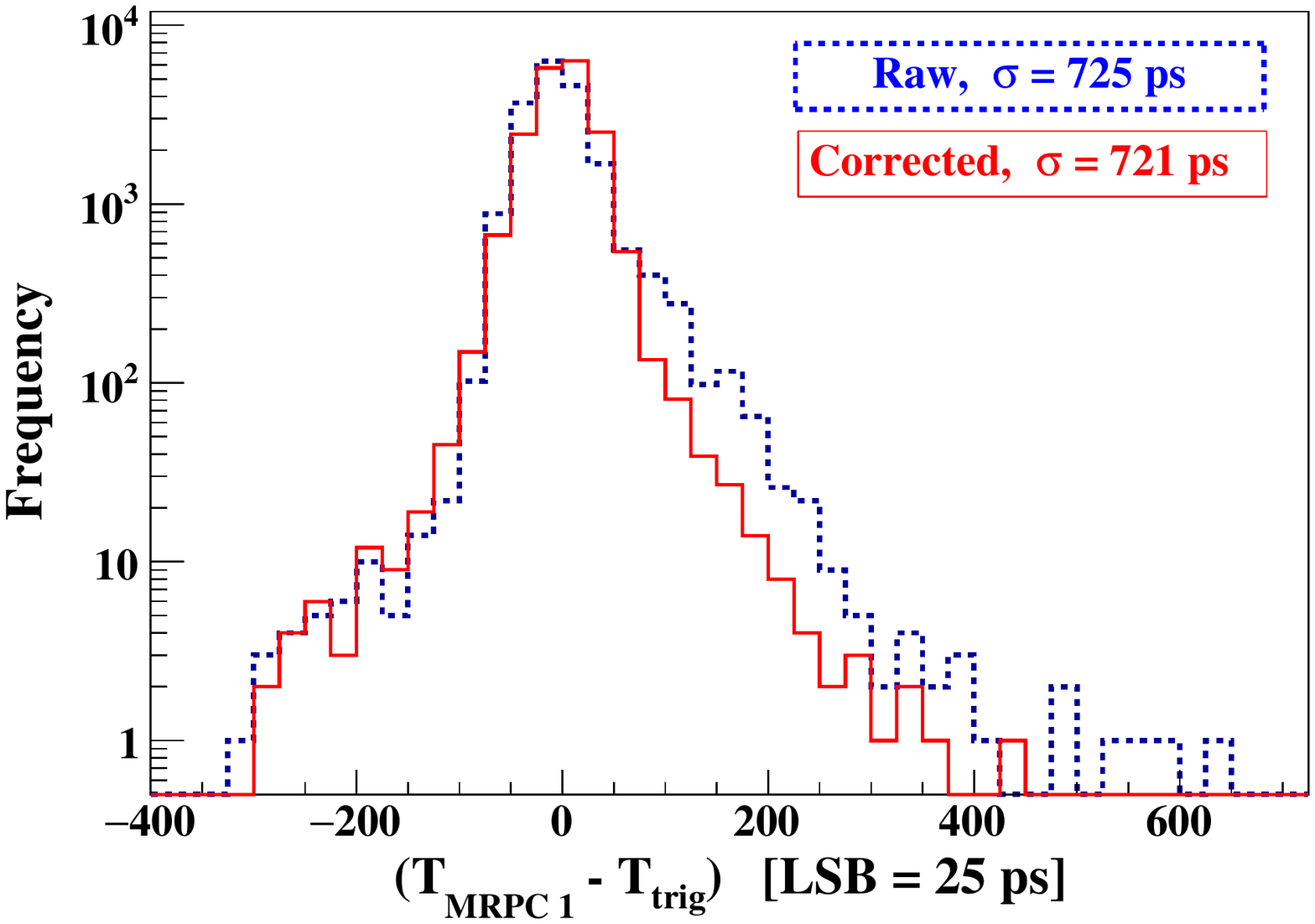}
\caption{The MRPC time distribution with respect to the trigger at 17.9 kV and 
with the gas mixture R134A ($91\%$), C$_4$H$_{10} (5\%)$
 and SF$_6 (4\%)$. The 
blue dashed line shows the raw distribution, while the red solid one shows the 
distribution after applying the time walk correction.  Note that the distributions have been 
shifted so that they peak at zero.}
\label{tdc-corr}
\end{figure}

The two dimensional profile histogram of time counts vs charge counts
is shown in Figure~\ref{calibration}. This is fitted to 
a function $\exp[- p_0/x + p_1 ] + p_2$. The time count of each event is then corrected according 
to the charge information 
by employing a calibration through the fit parameters as follows, 
\begin{equation}
 \rm T_{corrected} = T_{raw} - T_{c}
\end{equation}
where, $T_{c}$ is the correction to be obtained from the time counts as a function of the charge as 
obtained from the calibration fit. The comparison of the raw and corrected time distributions is shown in 
Fig.~\ref{tdc-corr}, which show clear reduction in the tail.
%\newpage

%%%%%%%%%%%%%%%%%%%%%%%%%%%%%%%%%%%%%%%%%%%%%%%%%%%%%%%%%%
%%%%%%%%%%%%%%%%%%%%%%%%%%%%%%%%%%%%%%%%%%%%%%%%%%%%%%%%%%
%%%%%%%%%%%%%%%%%%%%%%%%%%%%%%%%%%%%%%%%%%%%%%%%%%%%%%%%%%
%%%%%%%%%%%%%%%%%%%%%%%%%%%%%%%%%%%%%%%%%%%%%%%%%%%%%%%%%%

\begin{figure}[ht]
\centering
\includegraphics[trim=0cm 0cm 0cm 0cm, width=10cm]{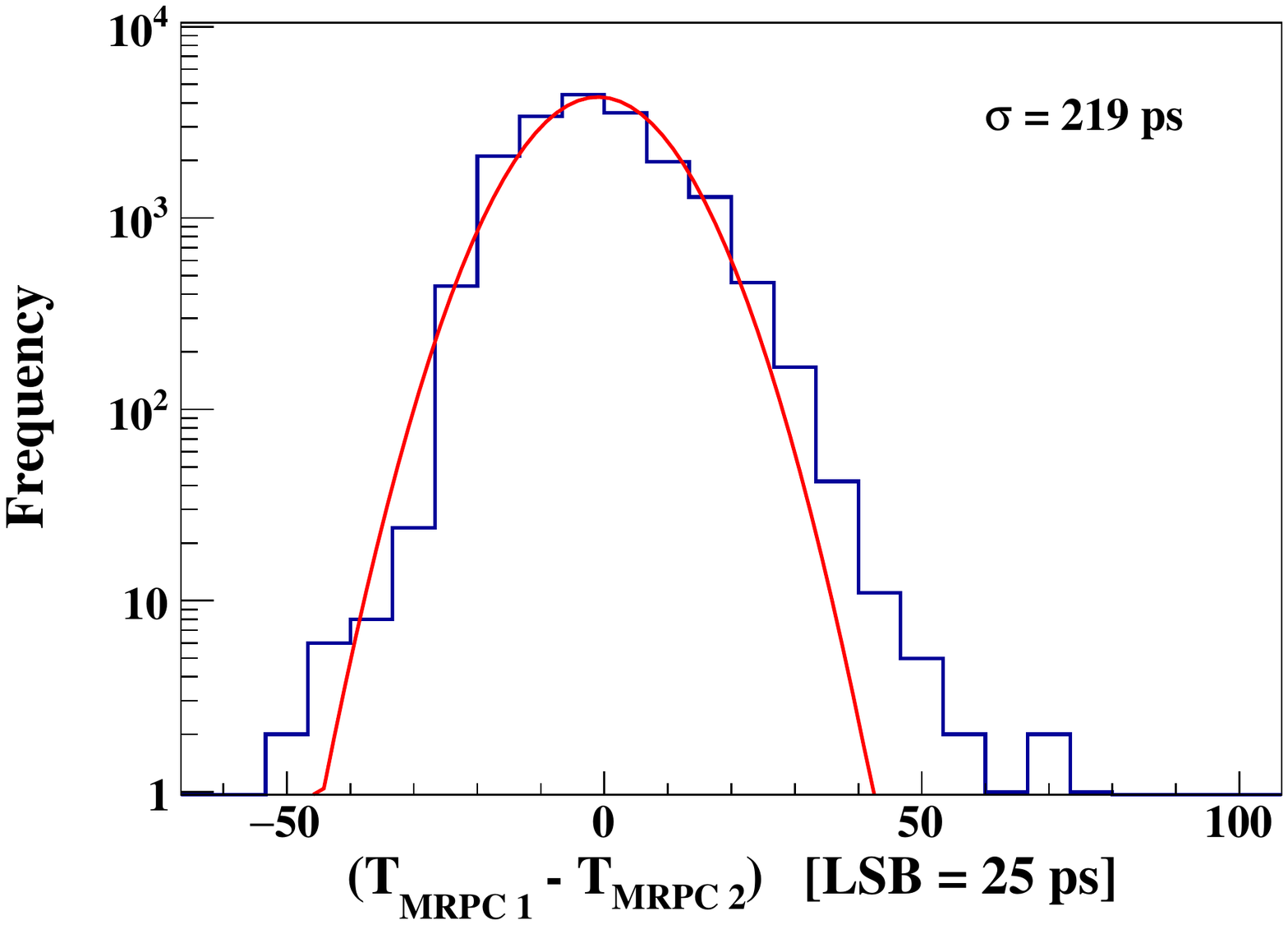}
\caption{an MRPC~1 time distribution with respect to MRPC~2 at 17.9 kV and 
with the gas mixture R134A ($91\%$), C$_4$H$_{10} (5\%)$
 and SF$_6 (4\%)$. The distribution is fitted to a Gaussian, and the time resolution, 
 i.e., the standard deviation of the fit is 219 ps.}
\label{tdc-corr1}
\end{figure}

The time distribution of MRPC1 with respect to MRPC2 is shown in Figure~\ref{tdc-corr1}, for $17.9$ kV. The time resolution is $219$ ps, 
which also includes $15$ -- $25$ ps of jitter from the electronics. This jitter 
has been estimated by replacing the MRPC strip signals with a pulser and 
observing the obtained time spectrum.

%%%%%%%%%%%%%%%%%%%%%%%%%%%%%%%%%%%%%%%%%%%%%%%%%%%%%%%%%%
%%%%%%%%%%%%%%%%%%%%%%%%%%%%%%%%%%%%%%%%%%%%%%%%%%%%%%%%%%

\subsection{MRPC as a part of trigger to single-gap RPC}
\label{srpc-test}

%%%%%%%%%%%%%%%%%%%%%%%%%%%%%%%%%%%%%%%%%%%%%%%%%%%%%%%%%%
%%%%%%%%%%%%%%%%%%%%%%%%%%%%%%%%%%%%%%%%%%%%%%%%%%%%%%%%%%
\begin{figure}[h]
\centering
\includegraphics[trim=0 15cm 0cm 3cm, width=12cm]{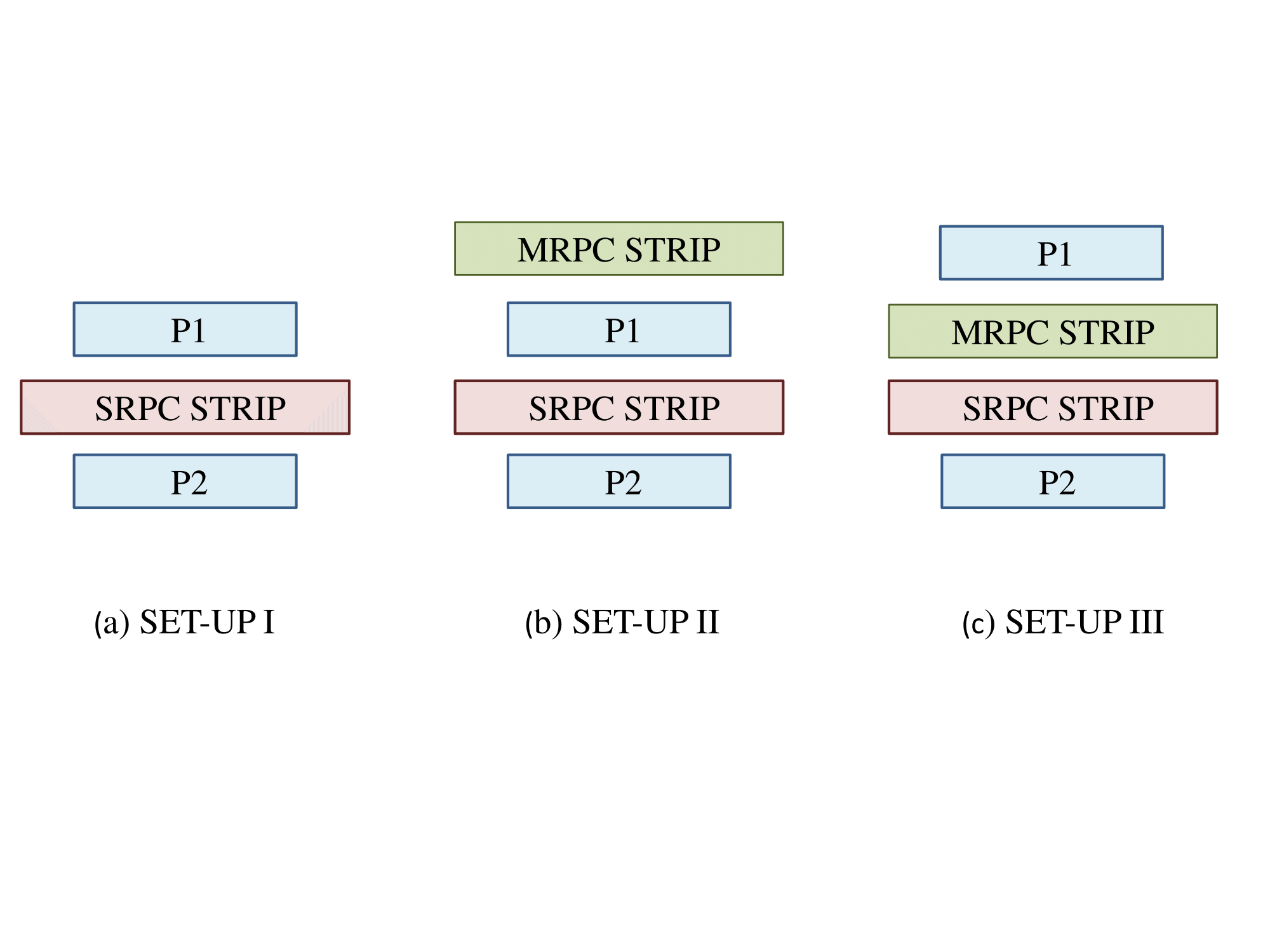}
\caption{The three set--ups under study. Set-up I consists only the scintillator paddles P1 and P2 in coincidence to 
form the trigger. The other two set--ups use an MRPC too.}
\label{trigger}
\end{figure}
%%%%%%%%%%%%%%%%%%%%%%%%%%%%%%%%%%%%%%%%%%%%%%%%%%%%%%%%%%

We then probe the MRPC potential by adding it to the external trigger system for a single-gap RPC. In Figure~\ref{trigger} 
we show the three different set--ups which were used. In the first set--up, the trigger is formed by 
 generating a coincidence between two scintillator paddles P$_1$ and P$_2$. For the other two set--ups, the MRPC 
 is added to the trigger. The various characteristics of the single-gap RPC from 
these three trigger set--ups are listed in Table~\ref{srpc-tb}. We see that the introduction of the MRPC in the trigger system 
helps in improving the time resolution. 

\vspace{1cm}

\begin{table}[ht]
\centering
\caption{The MRPC characteristics for different trigger set--up}
\begin{tabular}{|llllll|}
\hline
Set--up&Trigger&Eff.&Time~res. & Noise & I \\
~& & ($\%$)&(ns)& ($\frac{Hz}{cm^{2}}$)& (nA)\\
 
\hline %\hline
I&P1, P2&85 &1.42 &1.5 & 305\\
%\hline
II&P1, P2, MRPC&85.9 & 0.87&2.85 & 312\\
%\hline
III&P1, P2, MRPC&87.8 &0.85 &1.87 &311 \\
\hline
%\hline
\end{tabular}
\label{srpc-tb}
\end{table}

%\newpage
\section{Summary}
\label{summary}
In this paper we have presented the development procedure of six--gap
MRPC detectors and their performance. The MRPC design has been 
optimized to ensure a uniform and proper gas flow through the sub--gaps.
The gas mixture used is R134a ($91\%$), $\rm C_{4}\rm H_{10} (5\%)$, $SF_{6} (4\%)$. The
characteristics like efficiency, time resolutions etc. were
studied at different operating voltages. We see a marked improvement 
in the time resolution after applying an off--line correction for time-walk.
At an operating voltage of $17.9$ kV, the time resolution is obtained to be $219$ ps, 
including the electronic jitter. An MRPC, used as a part of the external trigger to a single gap RPC, 
also reduces the time jitter of the trigger significantly.  A stack of three MRPC detectors is now 
fully operational.

This setup will now be used for a TOF measurement study. It is also being planned to fabricate 
MRPCs with double cell configuration, which would enable us to explore the characteristics at 
applied voltages higher than $18$ kV. 

\section{Acknowledgement}
\label{acknow} 
This work is a part of the extended detector studies by the INO Collaboration. 
We would like to thank P. Verma, S. R. Joshi, S. Kalmani, Mandar
Saraf, Darshana Gonji, Santosh Chavan and Vishal
Asgolkar for their help during the course of this work. We also 
express our gratitude to Prof. V.M.~Datar and Prof. G.~Majumder 
for the critical reading of the manuscript and many useful suggestions. MMD acknowledges the 
support from the Department of Atomic Energy (DAE) and the Department 
of Science and Technology (DST), Government of India. MMD also thanks 
the Weizmann Institute of Science for their hospitality, where parts of the 
manuscript were written.

\end{document}